\documentclass[
 reprint,
 amsmath,amssymb,
 aps, 
 prl,
 english,
 nofootinbib
]{revtex4-2}

\usepackage{graphicx}
\usepackage{dcolumn}
\usepackage{bm}
\usepackage{physics}
\usepackage{enumerate}
\usepackage[hidelinks]{hyperref}
\usepackage{tikz}
\usepackage{xcolor}

\renewcommand{\grad}{\nabla}

\usepackage{babel}
\usepackage{comment}

\definecolor{mygreen}{rgb}{0.0,0.55,0.3}

\definecolor{myred}{rgb}{0.75,0.0,0.0}

\begin{document}
\preprint{APS/123-QED}
\title{Nonreciprocal Disorder Prevents Zero-Temperature Freezing in a Ferromagnet}

\author{Noah Grodzinski$^{1}$}\thanks{Corresponding author: njbg2@cam.ac.uk (he/him)}
\author{Robert L. Jack$^{1,2}$}
\author{Sarah A.M. Loos$^{3}$}

\affiliation{$^1$Department of Applied Mathematics and Theoretical Physics, University of Cambridge, Wilberforce Road, Cambridge, United Kingdom}
\affiliation{$^2$ Yusuf Hamied Department of Chemistry, University of Cambridge, Lensfield Road, Cambridge, United Kingdom}
\affiliation{$^3$ Max Planck Institute for Dynamics and Self-Organization, G\"ottingen, Germany}
\date{\today}
\begin{abstract}
    Nonreciprocal interactions underpin diverse nonequilibrium phenomena, yet the effects of quenched nonreciprocity in extended systems remain largely unexplored. We study a $2d$ Ising model with randomly distributed nonreciprocal bonds at density $p$, finding a continuous nonequilibrium transition down to $T=0$ with finite $p_c$. A gauge-invariance argument yields $p_c(T)\leq1/2$, and mean-field theory predicts a qualitatively correct phase diagram. Unlike equilibrium disordered models, the zero-temperature dynamics remains active, with athermal rare-region reversals and logarithmic ``activated'' coarsening.

\end{abstract}

\maketitle

\textit{Introduction--} Many collective systems are largely cooperative, yet a sparse minority of antagonistic, asymmetric interactions can have dramatic consequences for global order. In voter models, for instance, a few contrarian individuals in an otherwise conformist population can prevent global consensus formation~\cite{Masuda_2013, Mobilia_Petersen_Redner_2007}. Such asymmetric, or \textit{nonreciprocal (NR)}, interactions are known to drive systems out of equilibrium~\cite{Loos_Klapp_2020}, producing cyclic currents and large-scale dynamical structure~\cite{fruchart2026nonreciprocalmanybodyphysics, You_Baskaran_Marchetti_2020, Fruchart_Hanai_Littlewood_Vitelli_2021}. Much of this understanding, however, rests on models with \textit{uniform} nonreciprocity, where NR interactions are introduced homogeneously or in a fully-connected geometry. Yet spatially disordered, or \textit{quenched}, nonreciprocity arises in many of the systems which motivate the study of NR interactions, including flocks~\cite{Yllanes_Leoni_Marchetti_2017,Toner_Guttenberg_Tu_2018}, neural networks~\cite{Sompolinsky_Kanter_1986, Parisi_1986}, and heterogeneous ecological communities~\cite{Galla_2006,Roy_2020,Ros_Roy_Biroli_Bunin_Turner_2023, Allesina_Tang_2012}. The generic consequences of spatially quenched nonreciprocity remain largely unexplored, even though conflicting local objectives between pairs might be expected to frustrate large-scale order.

To study such questions theoretically, spin systems have proven a fertile minimal setting~\cite{fruchart2026nonreciprocalmanybodyphysics}. Models with two species of spins interacting in a uniformly nonreciprocal manner exhibit transitions from static order to oscillating states~\cite{Avni_Fruchart_Martin_Seara_Vitelli_2025a,Avni_Fruchart_Martin_Seara_Vitelli_2025b,Garcia_Lorenzana_Altieri_Biroli_Fruchart_Vitelli_2025,Blom_Thiele_Godec_2025,Guislain_Bertin_2024b,Guislain_Bertin_2024,Arjun_Kumar_2026}. Single-species models with spatially fixed, ordered NR couplings have been shown to exhibit modified phase transitions~\cite{Godreche_Luck_2017,Godreche_Bray_2009,Sanchez_Lopez_Rodriguez_2002, Lipowski_2015,Rajeev_Kumar_2024} and advection~\cite{Seara_Piya_Tabatabai_2023, Weiderpass_2025, Godreche_2011, DiCarlo_2025}, while configuration-dependent (e.g., vision-cone) interactions can stabilise long-range order and reshape defect dynamics via self-advection~\cite{Loos_Klapp_Martynec_2023, Bandini_Venturelli_Loos_Jelic_Gambassi_2025, Dopierala_etal_2025, Popli_Maitra_Ramaswamy_2025, Liu_Zheng_Nian_Xiong_2025, Rouzaire_Pearce_Pagonabarraga_Levis_2025,Garces_Levis_2025,dadhichi2020nonmutual}. Disordered NR interactions have, so far, been studied mainly in fully-connected (mean-field) models, where they typically destabilise ordered and spin-glass states in favour of persistent dynamics~\cite{Sompolinsky_Crisanti_Sommers_1988,Crisanti_Sompolinsky_1987,Crisanti_Sompolinsky_1988,Guislain_Bertin_2024d,Guislain_Bertin_2024c, Garcia_Lorenzana_Altieri_Biroli_Fruchart_Vitelli_2025}. Yet finite-dimensional glassy states can survive some NR driving~\cite{Marinari_Stariolo_1998,Berthier_Barrat_Kurchan_2000, klamser2025directedpercolationtransitionactive}, and ``nonreciprocal frustration'' can even lead to glassy ageing where it would not otherwise occur~\cite{Hanai_2024}.

A central, yet little studied, question is therefore: how does quenched nonreciprocity affect ordering in spatially extended systems? In equilibrium, the competition between quenched disorder and ferromagnetism has been the subject of decades of research: models such as the random-field and random-bond Ising models (RFIM~\cite{Imry_Ma_1975} and RBIM~\cite{Nishimori_2001, Binder_Young_1986, Nishimori_1981}) support a rich phenomenology including rare-region effects~\cite{Vojta_2006}, pinning~\cite{Huse_Henley_1985}, and zero-temperature freezing~\cite{Jain_1999b}. Here we seek to understand how these phenomena are modified when the quenched disorder is \textit{non}reciprocal.

To address this, we introduce a minimal nonreciprocal analogue of the $\pm J$ RBIM: a $2d$ Ising model with NR bonds inserted randomly at density $p$. We characterise the phases of this system at finite and zero temperature, establishing two main results. First, quenched nonreciprocity destroys ferromagnetic order: our model exhibits a continuous transition to a paramagnetic phase, with a critical temperature $T_c(p)$ that decreases to zero at finite $p_c$. Using a gauge-invariance argument, we show that long-ranged order requires percolation of the reciprocal bonds, giving $p_c \leq 1/2$. Second, and perhaps more surprisingly, we demonstrate that localised nonreciprocal driving induces persistent dynamics at zero temperature: freezing is prevented in the disordered phase, and behaviour typically associated with thermal noise (rare-region reversals, activated coarsening) persists down to $T=0$.

\begin{figure*}[t]
    \includegraphics[width = 1.0\linewidth]{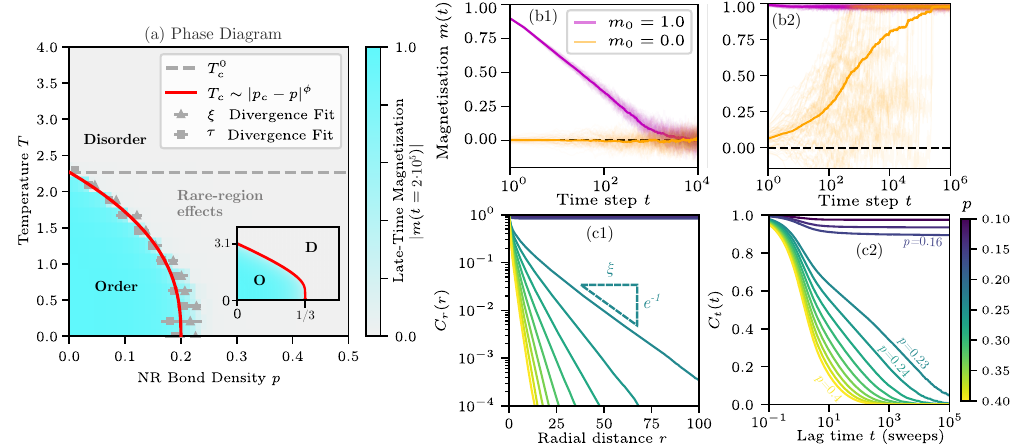}
    \caption{\label{fig:phases} Phases. (a) Phase diagram in the $(p, T)$-plane. Colours show late-time magnetisation ($t=2\times 10^5$, $N=512$, initialised at $m_0=1.0$; the positive $m_0$ leads to some residual magnetisation in the disordered phase near criticality). Squares and triangles mark $p_c(T)$ extrapolated from the divergences of $\xi$ and $\tau$ (see Supplemental Material); red curve is a fit $T_c \propto \abs{p_c-p}^\phi$. Inset: mean-field magnetisation and transition line, from Eq.~\eqref{eq:mf_finiteT}. (b) Relaxation at $T=0$ from magnetised (purple) and paramagnetic (orange) initial conditions, 64 disorder realisations with mean shown by darker, thick line: (b1) $p=0.3$, $N=128$; (b2) $p=0.1$, $N=32$ (small lattice allows full coarsening in accessible time, see \textit{Coarsening}). (c) Correlation functions at $T=0$ ($N=512$, averaged over $t \in [10^5, 10^6]$ and 64 realisations), shown only for runs that reached steady state. (c1) Spatial correlation (exponential decay) for various $p$; (c2) Autocorrelation. }
\end{figure*}

\textit{Model--} We consider Ising spins $\sigma_i \in \{-1,+1\}$ on an $N \times N$ square lattice with periodic boundaries. Each spin $i$ has a \textit{selfish energy}~\cite{Avni_Fruchart_Martin_Seara_Vitelli_2025b}
\begin{equation}\label{eq:selfish_energy}
    E_i = -\sum_{j \in \partial i} J_{ij}\,\sigma_i \sigma_j\,,
\end{equation}
where the sum runs over nearest neighbours. Couplings are drawn independently: for neighbours $\ev{i, j}$ we take
\begin{equation}
    (J_{ij}, J_{ji}) =
    \begin{cases}
        (+1, +1) & \text{with probability } 1-p, \\
        (+1, -1) & \text{with probability } p/2, \\
        (-1, +1) & \text{with probability } p/2.
    \end{cases}
    \label{eq:couplings}
\end{equation}
(Note: if $J_{ij}=-1$ then necessarily $J_{ji}=+1$.) We let each spin $i$ respond to its local field $h_i$ from its neighbours $j$, adopting Glauber dynamics~\cite{Glauber_1963} as in other NR spin models~\cite{ Loos_Klapp_Martynec_2023,Avni_Fruchart_Martin_Seara_Vitelli_2025a,fruchart2026nonreciprocalmanybodyphysics}:
\begin{equation}\label{eq:glauber}
    P_{\sigma_i \to -\sigma_i} = \frac{1 - \sigma_i\tanh\!\left(h_i/T\right)}{2}, \quad h_i = \sum_j J_{ij}\sigma_j,
\end{equation}
with $\Delta E_i = 2\sigma_i h_i$ the cost of flipping. At $T=0$, spins with $\Delta E_i < 0$ flip deterministically, those with $\Delta E_i > 0$ do not flip, and those with $\Delta E_i = 0$ flip with probability $1/2$. We define time $t$ such that $\Delta t = 1$ corresponds to $N^2$ randomly attempted flips.

The model shares features with the $\pm J$ RBIM~\cite{Nishimori_1981}: both have $\mathbb{Z}_2$ symmetry, and both draw quenched couplings from a discrete set $J_{ij}\in\{+1,-1\}$ with positive mean, instead of a continuous distribution. However, two distinctions from RBIM lead to important differences at low temperature. First, the dynamics cannot be described by a global free energy, permitting directed cycles in configuration space. Second, the disorder is \textit{directional}: each NR bond carries an orientation.

\textit{Order/Disorder Transition--} We first discuss which phases occur in our system. A suitable order parameter is the magnetisation $ m(t) = N^{-2} \sum_i \sigma_i(t), $ with mean steady-state value $M \equiv \lim_{t \to \infty} \overline{\langle |m(t)| \rangle}$. Here, $\langle\cdot\rangle$ denotes an average over realisations of the stochastic dynamics, and $\overline{\cdot}$ an average over quenched disorder.

We build a qualitative intuition for the phase diagram via a mean-field treatment, treating explicitly the five possible values of the local field on each site (see End Matter). At a single-site mean-field level, the local field distribution of our model maps onto the $\pm J$ RBIM: each coupling $J_{ij}$ entering the local field \eqref{eq:glauber} is independently $-1$ with probability $p/2$, and $+1$ otherwise. Solving the mean-field self-consistency equation gives ordered ($M > 0$) and disordered ($M = 0$) phases, separated by a second-order phase transition at $T_c(p)$ which decreases monotonically to zero at $p_c^{\mathrm{MF}} = 1/3$ [Fig.~\ref{fig:phases}(a, inset)].

The correspondence with the random-bond Ising models extends to the coarse-grained field level. Block-averaging the microscopic dynamics near criticality (see End Matter) yields a minimal field theory, in which nonreciprocity enters as a single \textit{randomly quenched advection} term on top of the existing terms from RBIM. We expect this kind of advection to be a generic consequence of quenched nonreciprocity~\cite{Seara_Piya_Tabatabai_2023} (see also~\cite{dadhichi2020nonmutual, Garces_Levis_2025}). Power-counting the extra term near criticality (or following~\cite{Lorenzana_Martin_Avni_Seara_Fruchart_Biroli_Vitelli_2025}) reveals that (i) nonreciprocity is \textit{irrelevant} near the clean-Ising fixed point (a Harris-like criterion $\nu(d+2)<2$ fails in all dimensions), and (ii) the \textit{symmetric} (dilution) part of the quenched disorder is always more relevant than the antisymmetric (NR) part near criticality, clarifying why our model appears similar to RBIM away from $T=0$.

Direct simulation of the Glauber dynamics on $N \times N$ lattices with periodic boundaries confirms this qualitative picture [see Fig.~\ref{fig:phases}(a)]. We find two phases, both independent of initial conditions [Fig.~\ref{fig:phases}(b)] (although the ordered steady state is only reached from a disordered initial condition in computationally accessible times for small system sizes -- see \textit{Coarsening}), separated by a continuous phase transition. We characterise these phases using spatial and temporal correlation functions
\begin{align}
    C_r(\vb{r}; t_0) & = \overline{\langle\sigma_i(t_0)\,\sigma_{i+\vb{r}}(t_0)\rangle}\,, \\
    C_t(t, t_0)      & = \overline{\langle\sigma_i(t_0)\,\sigma_i(t_0+t)\rangle}\, ,
\end{align}
writing $C_t(t)$ for the steady-state autocorrelation function ($t_0\to\infty$) and $C_r(r)$ for the radially averaged steady-state spatial correlation function.

\begin{figure}[t]
    \includegraphics[width = 1.0\linewidth]{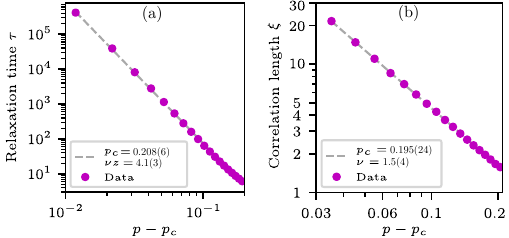}
    \caption{\label{fig:cont_transition} Evidence for a continuous zero-temperature phase transition, using correlations [shown in Fig.~\ref{fig:phases}(c)]. (a) Divergence of bulk relaxation time $\tau$, defined by $C_t(\tau)=1/e$, as $p \to p_c$. Dashed line is a fit to $\tau \sim |p-p_c|^{-\nu z}$. (b) Divergence of spatial correlation length $\xi$, fit to $\xi \sim |p-p_c|^{-\nu}$. Errors come mainly from varying the range of $p$ included in the fit. }
\end{figure}

In the ordered phase $T < T_c(p)$, $C_r(r)$ and $C_t(t)$ both approach a finite plateau at large distances and times, signalling long-range order. In the disordered phase, both correlation functions decay to zero at large $r$ as $C_r(r) \sim e^{-r/\xi}$, defining a finite spatial correlation length $\xi$, and $C_t(t)$ decays to $1/e$ on a finite timescale $\tau$ (although with stretched-exponential tails, see \textit{Rare-region effects}) [see Fig.~\ref{fig:phases}(c)]. Approaching the transition from the disordered side, both $\xi$ and $\tau$ diverge, including at zero temperature (Fig.~\ref{fig:cont_transition}), consistent with a continuous phase transition. Fitting these divergences to power laws in $|p - p_c|$ yields $p_c \simeq 0.20$, and allows us to numerically estimate the transition line $T_c(p)$ shown in Fig.~\ref{fig:phases}(a). This has the same qualitative features as the MF prediction, namely (i) the existence of a zero-temperature phase transition at finite $p_c$, and (ii) a decreasing $T_c(p)$ with no \textit{re-entrant disorder} (unlike in RBIM \cite{Nishimori_1981,ParisenToldin_Pelissetto_Vicari_2009}).

The picture above is constrained by an exact upper bound $p_c \leq 1/2$, established by a gauge-symmetry argument that we explain in the End Matter. When reciprocal bonds fail to percolate the lattice, a cluster gauge transformation maps each disorder realisation onto an equally probable one with the same dynamics but opposite sign of inter-cluster correlations, forcing $\overline{ \ev{\sigma_i \sigma_j} } = 0$ for sites in different clusters. Using the $2d$ bond percolation threshold, this gives $p_c(T) \leq 1/2$ for all $T \geq 0$. This argument relies on the fact that NR bonds have $J_{ij} = -J_{ji}$; no equivalent argument exists for equilibrium systems.

We now turn to the zero temperature dynamics, where the consequences of nonreciprocity are most pronounced: unlike RBIM~\cite{Gandolfi_Newman_Stein_2000,Jain_1999b}, the disordered phase does not freeze at $T=0$, already visible as vanishing autocorrelations in Fig.~\ref{fig:phases}(c). We first characterise this prevented freezing by examining the steady-state autocorrelation tails, before identifying its origin.

\begin{figure}[t]
    \includegraphics[width = 1.0\linewidth]{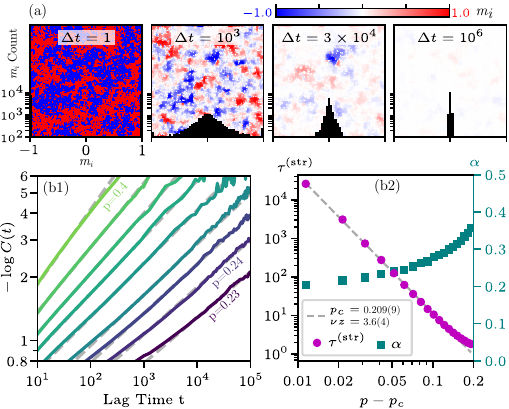}
    \caption{\label{fig:rare_regions} Rare region effects. (a) Steady-state configurations at $p\!=\!0.3$, $T\!=\!0$, $N\!=\!128$, averaged over increasing time windows (left to right): snapshot; average over bulk relaxation time; longer-time average where rare regions remain visible; average beyond the rare-region flipping time. Inset: histograms of time-averaged spins $m_i=\ev{\sigma_i}_{\Delta t}$ (log scale); heavy tails and slow convergence to $\delta(x)$ signal rare regions. (b1) Autocorrelation tails at varying $p$, on axes where stretched-exponential fits to Eq.~\eqref{eq:stretched} (grey dashes) are straight lines [same data as Fig.~\ref{fig:phases}(c2)]. (b2) Divergence of $\tau^\mathrm{(str)}$ towards $p_c$ and plateauing stretching exponent $\alpha$; dashed line is a fit to $\tau^\mathrm{(str)} \sim |p-p_c|^{-\nu z}$.
    }
\end{figure}

\textit{Rare-region effects--} The correlation functions of the low-temperature disordered phase are affected by the presence of rare ordered regions. Specifically, the \textit{tails} of the steady-state autocorrelation $C_t(t)$ exhibit a stretched-exponential decay~\cite{Bray1988},
\begin{equation}\label{eq:stretched}
    C_t(t) \sim \exp[-(t/\tau^\mathrm{(str)})^\alpha],
\end{equation}
with $\alpha < 1$, shown in Fig.~\ref{fig:rare_regions}(b) at $T=0$. These tails are a consequence of rare regions with locally lower NR bond density; such regions can sustain transient local order, and flip on a much larger timescale than the rest of the disordered bulk (see Supplemental Material for a derivation of \eqref{eq:stretched} and a discussion of the behaviour of $\alpha$). This is illustrated in Fig.~\ref{fig:rare_regions}(a, centre-right) -- rare regions remain visible when averaging over a time window much longer than the bulk relaxation time. This effect dominates the tails of the autocorrelation function, despite the sparsity of rare regions. Importantly, even though the same effect appears in RBIM~\cite{Bray1988}, rare-region reversals here persist at finite rate down to $T=0$ (as in Fig.~\ref{fig:rare_regions}), unlike in equilibrium models~\cite{Vojta_2006}.

\textit{Nonreciprocal driving--} We now discuss how nonreciprocity prevents freezing at zero temperature. In RBIM at $T=0$, a finite fraction of spins stop flipping entirely (``frozen spins''), producing a spin-glass phase~\cite{Gandolfi_Newman_Stein_2000,Newman_Stein_1999,Jain_Flynn_2006}. In contrast, \textit{all} sites in the disordered phase of our model remain dynamically active at zero temperature [see End Matter, Fig.~\ref{fig:EM_selfish_energy}(a,b)].

This is possible due to an asymmetry between the local and global energy, reminiscent of an Escher staircase~\cite{schuttler2025nonreciprocaldynamicsweaknoise}. In a reciprocal system, changes in the local selfish energy [Eq.~\eqref{eq:selfish_energy}] and global energy ($E=\sum_i E_i$) due to a spin flip are exactly proportional: $\Delta E = 2 \Delta E_i$. Since $E$ is single-valued and minimised in steady state, every permitted $T=0$ flip in a steady-state reciprocal system leaves the total energy unchanged: allowed flips are ``agnostic'' (e.g., zero-local-field ``blinker'' sites~\cite{Olejarz_Krapivsky_Redner_2011}). This relationship is violated by nonreciprocity -- a flip can \textit{decrease} the local selfish energy $E_i$, and so occur deterministically, while increasing or leaving unchanged the global $E$ (specifically, $\abs{\Delta E - 2\Delta E_i} \leq 4n_\text{NR}$ for $n_\text{NR}$ nonreciprocally coupled neighbours of $i$). The local selfish energy can therefore decrease indefinitely while the global one remains constant.

Numerically, a large fraction of $T=0$ steady-state flips have $\Delta E_i < 0$ [see End Matter, Fig.~\ref{fig:EM_selfish_energy}(c, d)]. A minimal configuration that allows such Escher-like infinite descent is a $2\times2$ plaquette of NR bonds with zero ``boundary field'' [Eq.~\eqref{eq:boundary-field}], which executes an 8-step deterministic cycle; an infinite family of such configurations exists (see End Matter). These deterministically flipping spins act as a source of localised driving, and appear to be dense enough to prevent freezing of \textit{any} spin in the disordered steady state. Behaviour that usually requires finite temperature (rare-region reversals and activated coarsening below) can then persist down to $T=0$.

\begin{figure}[t]
    \includegraphics[width = 1.0\linewidth]{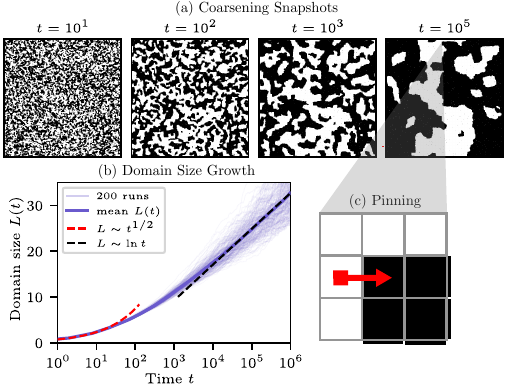}
    \caption{\label{fig:pinning} Activated coarsening in the ordered phase. (a)~Snapshots after a quench from $T=\infty$ ($512 \times 512$, $p=0.05$). (b)~Domain size defined by $C_r(L) = 1/e$; 200 disorder realisations ($400 \times 400$, $p=0.1$) (light), mean (dark): Early-time growth $L \sim t^{1/2}$ (red dashes show fit) crossing to $L \sim \ln t$ (black), around $L^* \simeq 10$. (c)~Pinning defect: colours show $\sigma_i$; arrow $i \to j$ marks NR bond $(J_{ij}, J_{ji}) = (+1,-1)$. }
\end{figure}

\textit{Coarsening--} Nonreciprocal driving prevents the disordered phase from freezing; we now show that it also facilitates coarsening in the ordered phase, by preventing ``pinning'' from arresting domain growth. Starting from a paramagnetic initial condition, the system orders through the usual coarsening of ferromagnetic domains. In RBIM, domain-wall pinning halts this process at $T=0$, leaving a permanently frozen domain configuration~\cite{Jain_1999}. Our model instead coarsens indefinitely, in two stages. At early times, domain walls move freely and $L(t) \sim t^{1/2}$ (standard curvature-driven growth~\cite{Bray_1994}). At later times, domain walls encounter directed NR bonds that act as pinning sites, and growth crosses over to an ``activated'' form, $L(t) \sim \ln t$ [Fig.~\ref{fig:pinning}(a,b)]~\cite{Huse_Henley_1985}. The pinning mechanism is illustrated in Fig.~\ref{fig:pinning}(c): a single NR bond can stabilise a corner or curved boundary of a domain, meaning the neighbouring domain walls must move for coarsening to proceed. Unlike RBIM, however, driving by NR interactions acts as a source of athermal dynamics; fully absorbing configurations at $T=0$ (e.g., a domain with a pinning defect at each corner and no NR driving along its walls) require coordinated bond arrangements along an extended boundary, becoming exponentially unlikely as the domain grows. As a result, activated growth can continue indefinitely at zero temperature in our model (at least for computationally accessible times), unlike the corresponding equilibrium process.

\textit{Discussion--} We have explored the effects of quenched nonreciprocity on ferromagnetic ordering by randomly inserting NR bonds at density $p$ into a $2d$ Ising model. We find that the finite-temperature phases are similar to their counterparts in the $\pm J$ RBIM, with ordered and disordered phases separated by a continuous transition. These similarities arise from an equivalence at the single-site mean-field level, and persist after coarse-graining. The zero-temperature behaviour, however, differs sharply: while RBIM is guaranteed to freeze, nonreciprocal driving in our model continues to drive dynamics. This means that behaviour typically associated with thermal noise, such as rare-region reversals and depinning of domain walls, persists down to $T=0$.

We expect these central features to be generic consequences of quenched nonreciprocity. This is in close analogy to fully-connected systems, like neural-network models~\cite{Crisanti_Sompolinsky_1987,Crisanti_Sompolinsky_1988,Sompolinsky_Crisanti_Sommers_1988} and disordered ecosystems~\cite{Galla_2006,Roy_2020,Ros_Roy_Biroli_Bunin_Turner_2023,Allesina_Tang_2012}, where asymmetry destroys fixed points and produces temporal chaos. Our results demonstrate that this effect persists in finite-dimensional ferromagnetic systems, and identify its origin: the gauge bound guarantees disorder once NR bonds are sufficiently dense, and nonreciprocity supplies a ``run-and-chase'' cyclic source of deterministic athermal driving which keeps the disordered phase active.

Several directions follow naturally. The most important is to test the generality of our results: in higher dimensions, analogous equilibrium models support a finite-temperature spin glass, and it is unclear under what conditions quenched nonreciprocity can destabilise this glassy order. Separately, a precise characterisation of critical exponents at the zero-temperature phase transition would situate this within a known or new universality class.

\textit{Acknowledgements---}We thank Michael Cates, Peter Sollich, Thomas Suchanek, Ludovic Berthier, and Filippo de Luca for insightful discussions throughout this work. This work was supported by Engineering and Physical Sciences Research Council under Grants Nos.~EP/W524633/1 (Project Reference 2927750) and EP/Z534766/1. This research was supported in part by grant NSF PHY-2309135 to the Kavli Institute for Theoretical Physics (KITP).

\vspace{-10pt}

\bibliography{apssamp}

\newpage
\section{END MATTER}
\textit{Gauge Symmetry and Upper Bounds on $p_c$-- } In disordered equilibrium spin models, symmetries of the disorder distribution constrain disorder-averaged quantities exactly~\cite{Nishimori_2001, Nishimori_1981}. We show that a similar argument applies to our stochastic dynamics.

Let $\mathcal{C}$ be a complete cluster of sites connected by reciprocal bonds ($J_{ij}=J_{ji}=+1$). Define the cluster gauge transformation (as in~\cite{Nishimori_1981})
\begin{equation}\label{eq:gauge_cluster}
    \mathcal{G}_\mathcal{C}:\quad \sigma_i\to\tau_i\sigma_i\,,\qquad
    J_{ij}\to\tau_i\tau_j\,J_{ij}\,,
\end{equation}
with $\tau_i=-1$ for $i\in\mathcal{C}$ and $\tau_i=+1$ otherwise.

\textit{Invariance of the disorder distribution.} Bonds with both endpoints in $\mathcal{C}$, or both outside, are unchanged ($\tau_i\tau_j=+1$). Bonds crossing the boundary of $\mathcal{C}$ are necessarily nonreciprocal (otherwise $\mathcal{C}$ would extend further), and $\tau_i\tau_j=-1$ maps $(J_{ij},J_{ji})\to(-J_{ij},-J_{ji})$, i.e., $(+1,-1)\leftrightarrow(-1,+1)$, which has equal probability. Hence $P(\{J\})$ is invariant under $\mathcal{G}_\mathcal{C}$.

\textit{Invariance of the dynamics.} The Glauber flip rate
\begin{equation}
    w_i(\sigma, J) = \frac{1}{2}\left[1 - \sigma_i\tanh\!\left(
        \frac{1}{T}\sum_{j\in\partial i}J_{ij}\sigma_j\right)\right]
\end{equation}
depends on $(\sigma,J)$ only through the products $J_{ij}\sigma_i\sigma_j$, each of which is invariant since $\tau_i\tau_j\cdot\tau_i \cdot \tau_j=1$. All transition rates are therefore invariant under $\mathcal{G}_\mathcal{C}$.

For every realisation $\{J\}$ containing a cluster $\mathcal{C}$, we can therefore construct an equally probable realisation $\mathcal{G}_\mathcal{C}(\{J\})$ with the same dynamics but for which $\langle\sigma_i\sigma_j\rangle$ has opposite sign whenever $i\in\mathcal{C}$, $j\notin\mathcal{C}$. These contributions cancel in the average over realisations of disorder:
\begin{equation}
    \overline{\langle\sigma_i\sigma_j\rangle}=0
    \quad\text{for } i\in\mathcal{C},\; j\notin\mathcal{C}.
\end{equation}
Note that for a \textit{fixed} realisation of $\{J\}$, the correlation need not vanish; cancellation is enforced through disorder averaging. The case $p=1$ is the special case in which every $\mathcal{C}$ is a single site, giving $\overline{\langle\sigma_i\sigma_j\rangle}=\delta_{ij}$.

Correlations can therefore survive the disorder average only between sites in the same reciprocal cluster. For $p>1/2$ reciprocal bonds do not percolate on the square lattice~\cite{Kesten_1980}, so all reciprocal clusters are almost surely finite and the probability of two sites at distance $r$ sharing a cluster decays exponentially in $r$. The disorder-averaged (or, equivalently, the site-averaged) correlation function inherits this decay, establishing
\begin{equation}
    p_c \leq \tfrac{1}{2}\,.
\end{equation}
This generalises to any lattice in any dimension polluted by totally NR bonds, giving a corresponding bound set by the respective bond percolation threshold: ferromagnetic order requires percolation of the reciprocal bond network.

\vspace{1em}
\textit{Mean-Field Analysis-- } We derive a discretised mean-field approximation for the phase diagram. The standard \textit{continuous} replacement $\sigma_j \to \langle\sigma_j\rangle = m$ for neighbours $j$ of a site $i$ is inadequate here: it gives $T_c(p)=4(1-p)$, predicting $T_c\to0$ only as $p\to1$, incompatible with the bound $p_c\leq1/2$ above. We instead set $\sigma_j=\pm1$ with \textit{mean} $\langle\sigma_j\rangle=m$, following~\cite{Honmura_Kaneyoshi_1979}, drawing each neighbour $\sigma_j$ of a site $i$ independently with $P(\sigma_j=\pm1)=(1\pm m)/2$.

Since each site $i$ sees anti-aligning $J_{ij}=-1$ with probability $p/2$, the single-neighbour field contribution $\varepsilon_j\equiv J_{ij}\sigma_j=\pm1$ has $P(\varepsilon_j=\pm1)=[1\pm m(1-p)]/2$, so the total local field is $h_i=\sum_{j=1}^4\varepsilon_j=2n-4$ with $n\sim\mathrm{Binom}(4,q)$ and
\begin{equation}
    q = \frac{1 + m(1-p)}{2}.
\end{equation}
Unlike the continuous approximation, we have retained a full local-field distribution. The single-spin dynamics has the stationary condition $\langle\sigma_i\rangle = \langle\tanh(h_i/T)\rangle$~\cite{Suzuki_Kubo_1968}. Evaluating this using the binomial distribution of $h_i$ gives
\begin{equation}\label{eq:mf_finiteT}
    m = \sum_{n=0}^{4}\binom{4}{n}q^n(1-q)^{4-n}
    \tanh\!\left(\frac{2n-4}{T}\right).
\end{equation}
Linearising around $m=0$ yields the critical condition
\begin{equation}\label{eq:mf_Tc_implicit}
    1 = \frac{1-p}{2}\left[\tanh\!\left(\frac{4}{T_c}\right)
        + 2\tanh\!\left(\frac{2}{T_c}\right)\right],
\end{equation}
shown in Fig.~\ref{fig:phases}(a, inset). The critical temperature decreases monotonically from $T_c(p=0)\approx3.09$ to zero at $p_c^\mathrm{MF}=1/3$, meeting the $T=0$ axis with infinite slope, $|dT_c/dp|\to\infty$.

\label{app:coarsegrain}
\vspace{1em}
\textit{Coarse-Grained Field Theory-- } We review the steps to derive a stochastic field theory for our model, using a mean-field gradient-expansion coarse-graining. We average the dynamics ($\partial_t\langle\sigma_i\rangle = -\langle\sigma_i\rangle + \langle\tanh(h_i/T)\rangle$ with $h_i=\sum_{j\in\partial i}J_{ij}\sigma_j$~\cite{Suzuki_Kubo_1968}) over blocks of size $\ell$ (with $1\ll\ell\ll\xi$, valid for large $\xi$, i.e., near criticality), such that the mean magnetisation field $m(\vb{x},t)=\ell^{-d}\sum_{i\in \vb{x}}\ev{\sigma_i}$ is slowly varying over space. We adopt the mean-field closure $\langle\tanh(h_i/T)\rangle \simeq \tanh(\langle h_i\rangle/T)$, reinserting the leading neglected fluctuation as thermal noise.

We split the couplings into symmetric and antisymmetric parts $J^{S/A}_{ij}=\tfrac12(J_{ij}\pm J_{ji})$. The contribution of each bond to the block-averaged local field (obtained by Taylor expanding $\ev{\sigma_i}$ about the midpoint of neighbouring sites $i$ and $j$ to first order in gradients) then becomes $J_{ij}\ev{\sigma_j} + J_{ji}\ev{\sigma_i} = 2J^S_{ij}\,m + J^A_{ij}\,{\mathbf{e}}_{ij}\!\cdot\!\nabla m + \mathcal{O}(\grad^2)$. For $p=0$, block-averaging recovers Model A~\cite{Hohenberg_Halperin_1977}, $\partial_t m = \nabla^2 m - V'(m) + \zeta$, with $V(m) = \tfrac{r}{2}\,m^2 + \tfrac{u}{4}\,m^4$ and a reduced-temperature or ``mass'' $r \propto(T-T_c)$. For $p>0$, the symmetric couplings $J^S$ form a \textit{diluted} ferromagnet whose density varies from block to block; these static fluctuations shift the local critical temperature
$
    r(\mathbf{x}) = r + \delta r(\mathbf{x}),
$
with uniform part $r=1 - 4(1-p)/T$ and fluctuation $\delta r(\mathbf{x})$~\cite{Grinstein_Luther_1976, Harris_1974}. The antisymmetric couplings $J^A$ locally break spatial-inversion (and isotropy) symmetry, and contribute a quenched, zero-mean \textit{random advective} term $\sim \mathbf{v}(\mathbf{x})\cdot\nabla m$ with $\mathbf{v}(\mathbf{x})\propto \sum_{\ev{i,j}\,\in\,\mathbf{x}} J^A_{ij}\,\vb{e}_{ij}$ (the sum of randomly orientated NR bonds in a block). Together:
\begin{equation}\label{eq:cg_eom}
    \partial_t m = \nabla^2 m - r(\mathbf{x})\,m - u\,m^3
    + \mathbf{v}(\mathbf{x})\!\cdot\!\nabla m + \zeta\,,
\end{equation}
with thermal noise $\zeta$ and quenched disorder
\begin{align}
    \overline{\delta r(\mathbf{x})} =0\,, \quad
    \overline{\delta r(\mathbf{x})\delta r(\mathbf{x}')}
     & =\Delta_r\,\delta(\mathbf{x}-\mathbf{x}')\,,   \label{eq:var_dr}   \\
    \overline{v_\alpha(\vb{x})} =0\,, \quad
    \overline{v_\alpha(\mathbf{x})v_\beta(\mathbf{x}')}
     & =\Delta_v\,\delta_{\alpha\beta}\,\delta(\mathbf{x}-\mathbf{x}')\,,
\end{align}
where $\Delta_r\propto p(1-p)$ and $\Delta_v\propto p$. [Retaining the full discrete local-field distribution instead of using the $\tanh$ closure above gives $r \propto 1-(1-p)(\tanh\tfrac2T+\tfrac12\tanh\tfrac4T)$, recovering \eqref{eq:mf_Tc_implicit} at $r=0$.] Nonreciprocity enters as a single new advective term on top of the existing RBIM field theory, consistent with earlier studies~\cite{Seara_Piya_Tabatabai_2023, Weiderpass_2025, fruchart2026nonreciprocalmanybodyphysics}. This term \textit{cannot} derive from a free energy, and represents the leading nonequilibrium effect of spatially quenched nonreciprocity.

We can power-count~\eqref{eq:cg_eom} to determine the relevance of the quenched disorder at the clean Ising fixed point (or equivalently, follow the MSRJD action procedure of Ref.~\cite{Lorenzana_Martin_Avni_Seara_Fruchart_Biroli_Vitelli_2025}). The reduced-temperature has $[r]=1/\nu$ (using $\xi\sim r^{-\nu}$ and $[\xi]\equiv-1$), so $[\delta r]=[r]=1/\nu$. The advective random velocity field couples to $\nabla m$, and $[\nabla]=1$, so $[\mathbf v]=[\delta r] - 1=1/\nu-1$. Using \eqref{eq:var_dr}:
\begin{equation}
    [\Delta_r]=\frac{2}{\nu}-d,\qquad [\Delta_v]=[\Delta_r]-2=\frac{2}{\nu}-d-2\,.
\end{equation}
Relevance of the dilution term recovers the Harris criterion $d\nu<2$~\cite{Harris_1974}, while nonreciprocity is relevant only for $\nu(d+2)<2$. Thus (i) random-advection is strictly less relevant than random-dilution near criticality, and (ii) quenched nonreciprocity is RG-irrelevant at the clean-Ising critical point.

\begin{figure}[t!]
    \includegraphics[width = 1.0\linewidth]{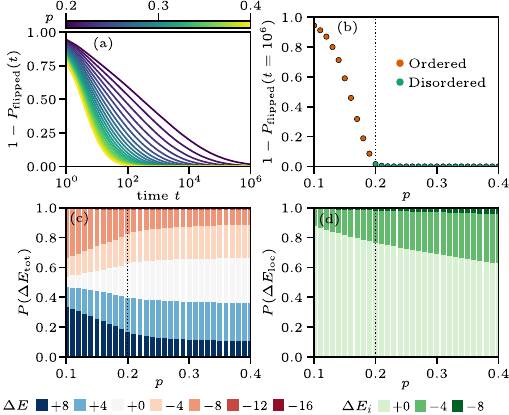}
    \caption{\label{fig:EM_selfish_energy} Prevented freezing by nonreciprocity. (a) Fraction of spins \textit{not} flipped up to time $t$ in the $T=0$ disordered state (averaged over 8 realisations of $512^2$ spins for each $p$, starting after $t_0=10^5$). (b) Proportion of spins which have not flipped between $t_0$ and $t_f=t_0+10^6$, same data. (c, d) Change due to a spin flip of the \textit{local} and \textit{global} energy at steady state; the average global energy change vanishes, while the local selfish energy changes are on average negative.}
\end{figure}

\vspace{1em}
\textit{Nonreciprocal Driving-- } In the disordered steady state, all sites remain dynamically active [Fig.~\ref{fig:EM_selfish_energy}(a,b)]. We attribute this activity to the presence of specific NR bond configurations, whose activity decreases the local selfish energy [Fig.~\ref{fig:EM_selfish_energy}(d)] while leaving the global energy unchanged [Fig.~\ref{fig:EM_selfish_energy}(c)]. We now demonstrate that a specific, infinite family of NR bond configurations has this property of persistent dynamics.

Consider a cluster $\mathcal{L}$ of sites $i$ whose internal bonds are all nonreciprocal. The field $h_i$ on a cluster spin [Eq.~\eqref{eq:glauber}] splits into an internal part from neighbours within $\mathcal{L}$ and a boundary field
\begin{align}\label{eq:boundary-field}
    h_i^{\partial} = \sum_{j \in \partial i \setminus \mathcal{L}} J_{ij}\sigma_j
\end{align}
from neighbours outside the cluster. Define the ``satisfaction'' $Q = \sum_{i\in\mathcal{L}} \sigma_i h_i $
i.e., minus the sum of selfish energies of the clustered spins. Since $J_{ij}=-J_{ji}$ on every NR bond, the contribution of internal bonds vanishes, so $Q = \sum_{i\in\mathcal{L}} \sigma_i h_i^{\partial}$ is only due to the boundary fields.

Consider first the case of vanishing boundary field, $h_i^{\partial}=0$ for all $i\in\mathcal{L}$. Then $Q=0$. A frozen (absorbing) state requires every spin strictly satisfied, $\sigma_i h_i \geq 1$, giving $Q\geq|\mathcal{L}|>0$ -- a contradiction. At least one spin can therefore always flip, so an all-NR cluster with vanishing boundary field cannot possess an absorbing state.

For $B$ lattice edges connecting a site in $\mathcal{L}$ to a site outside $\mathcal{L}$, the maximum possible contribution to $Q$ of the external fields is bounded by $B$, so $Q\leq B$. A frozen state still requires $Q\geq|\mathcal{L}|$. Hence any cluster with $|\mathcal{L}|>B$ admits no stationary state, irrespective of $h^\partial$.

Finally, every cluster with $\abs{\mathcal{L}}\geq2$ supports a directed cycle. Let $V^\ast$ be the set of cluster configurations with at least one nonzero field $h_i$. This set is nonempty for $\abs{\mathcal{L}}\geq 2$: fixing all spins but one neighbour $b$ of a site $a$, the two choices of $\sigma_b$ shift $h_a$ by $\pm2$, so cannot both give $h_a=0$. Every $\sigma\in V^\ast$ admits a deterministic move, since $Q=0$ with not all terms zero forces some site to have $\sigma_i h_i\leq-1$. And such moves remain in $V^\ast$: flipping a spin leaves its own field unchanged, so flipping a site with $h_i\neq0$ cannot reach an all-$h_i=0$ configuration. A deterministic trajectory confined to the finite set $V^\ast$ must revisit a configuration, therefore closing into a directed cycle.

\clearpage
\onecolumngrid
\graphicspath{{}}

\begin{center}
    {\large\bfseries Supplemental Material}\\[0.5em]
\end{center}
\vspace{1.5em}

\setcounter{section}{0}
\setcounter{subsection}{0}
\setcounter{equation}{0}
\setcounter{figure}{0}
\setcounter{table}{0}
\renewcommand{\thesection}{S\arabic{section}}
\renewcommand{\thesubsection}{\thesection\Alph{subsection}}
\renewcommand{\theequation}{S\arabic{equation}}
\renewcommand{\thefigure}{S\arabic{figure}}
\renewcommand{\thetable}{S\arabic{table}}

\setlength{\textfloatsep}{8pt plus 2pt minus 2pt}
\setlength{\floatsep}{8pt plus 2pt minus 2pt}

This Supplemental Material contains two sections. First we detail the procedure used to extract the finite-temperature transition line $T_c(p)$ [equivalently $p_c(T)$] shown in Fig.~1 of the main text; we then discuss a finite-time scaling collapse procedure used to extract the zero-temperature critical exponents. The second section develops the nonequilibrium rare-region theory of autocorrelation functions in the dynamically disordered phase, following the arguments of~\cite{Bray1988}, discussed in the \textit{Rare-region effects} section of the main text.

\section{Extracting the finite-temperature transition line $T_c(p)$}
\label{sm:transition}

\subsection{Data generation}

The simulations are performed on a $2d$ square lattice of $N\times N = 512\times512$ Ising spins with periodic boundary conditions, using the generalised Glauber dynamics as defined in the main text [Eqs.~(1--3)]. Each directed bond $(i\to j)$ carries its own coupling $J_{ij}$: reciprocal pairs have $J_{ij}=J_{ji}=+1$, while a fraction $p$ of bonds are made nonreciprocal (NR) by introducing an opposing (antiferromagnetic) coupling in a single direction: $J_{ij}=-1$, $J_{ji}=+1$, or vice versa, each with probability $p/2$. The density $p$ interpolates between the reciprocal Ising model ($p=0$) and a fully nonreciprocal lattice ($p=1$).

Dynamics are single-spin-flip Glauber updates at temperature $T$, with one sweep ($\Delta t = 1$) defined as $N^2$ attempted flips on randomly chosen sites (such that the expectation of the number of times each site can flip in a given sweep is $1$). Each $(p,T)$ point is run for $2\times10^5$ sweeps starting from the uniformly polarised state $m(0)=1$. Initialising in the ordered state lets us identify the \textit{disordered} phase by measuring demagnetisation. The magnetisation $m(t)$ is recorded every $10$ sweeps ($2\times10^4$ samples per trajectory). After an initial transient, the (non-connected) steady-state spatial correlator
\begin{equation}
    C_r^{\mathrm{nonn}}(r) = \overline{\langle \sigma_i\,\sigma_{i+r}\rangle},
\end{equation}
and the spin autocorrelation
\begin{equation}
    C_t(t) = \overline{\langle \sigma_i(t_0)\,\sigma_i(t_0+t)\rangle},
\end{equation}
are accumulated over the second half of the trajectory; $C_t$ is reported on a grid of lags up to $10^5$ sweeps, spaced by $10$ sweeps. Here, $\langle\cdot\rangle$ averages over the stochastic dynamics and $\overline{\cdot}$ over disorder, as in the main text. The superscript ``nonn'' denotes the non-normalised, non-connected correlator before subtraction of $\overline{\langle m\rangle}^2$.

The grid covers $p\in\{0,0.01,\ldots,0.40\}$ ($41$ uniformly spaced values) and $T\in\{0,0.209,\ldots,2.30\}$ ($12$ values, uniformly spaced between $0$ and slightly above the $2d$ Ising critical temperature $T_c^{\text{Ising}}\approx2.269$). One long trajectory is run per $(p,T)$ point, with the run length providing the statistical averaging for the time- and space-resolved correlators.

\subsection{Transition line extraction procedure}

The procedure below converts the per-run $C_r^{\mathrm{nonn}}$ and $C_t$ into the transition line $p_c(T)$ and the exponents $\nu(T)$, $\nu z(T)$.

\paragraph{Disordered onset.} A run is classified as disordered if $\overline{\langle|m(t)|\rangle}$, averaged over the final $25\%$ of the run, falls below $0.1$. At fixed $T$, the smallest $p$ satisfying this criterion defines $p_{\mathrm{dem}}(T)$; only points with $p>p_{\mathrm{dem}}$ enter the power-law fits, since the fits target the approach to criticality \emph{from the disordered side}.

\paragraph{Correlation length $\xi$.} We define the connected, normalised spatial correlator
\begin{equation}
    C_r(r) \equiv \frac{C_r^{\mathrm{nonn}}(r) - \overline{\langle m\rangle}^{\,2}}
    {C_r^{\mathrm{nonn}}(0)},
\end{equation}
and extract $\xi$ from an exponential fit
\begin{equation}\label{eq:sm_xi_fit}
    C_r(r) = A\,e^{-r/\xi}
\end{equation}
on the window $\{r>5\}\cap\{C_r>10^{-4}\}\cap(\{r<15\}\cup\{C_r>10^{-2}\})$. The lower bound on $r$ excludes the short-distance, non-universal regime; the upper bound is relaxed where the data retain high signal-to-noise.

\paragraph{Correlation time $\tau$.} We take $\tau$ to be the lag at which the normalised autocorrelation $C_t(t)/C_t(0)$ first crosses $1/e$, with linear interpolation between sampled lags. This threshold-crossing definition is robust to the noisy long-time tail of $C_t$ near criticality. Values $\tau<10$ are discarded as unresolved, since we only sample $C_t$ once per $10$ sweeps.

\paragraph{Power-law fits.} On the disordered side, we fit $\xi$ and $\tau$ to divergences of the form
\begin{equation}\label{eq:sm_powerlaws}
    \xi(p) = A_\xi\,|p-p_c|^{-\nu}, \qquad
    \tau(p) = A_\tau\,|p-p_c|^{-\nu z}.
\end{equation}
At each $T$, we fit a narrow band of points just above the disordered onset, $p_{\mathrm{dem}}<p<\max(2p_{\mathrm{dem}},\,p_{\mathrm{dem}}+0.1)$ for $\xi$ and $p_{\mathrm{dem}}<p<\max(1.5p_{\mathrm{dem}},\,p_{\mathrm{dem}}+0.1)$ for $\tau$. Restricting to this band keeps the fit within the asymptotic scaling window and away from corrections to scaling at large $|p-p_c|$. We restrict the fits to the disordered side because the ordered-side data are both sparser and noisier: the connected correlators decay over a smaller range before plateauing, reducing the accuracy of the extracted $\xi$ and $\tau$. The slice $T=2.30$ is excluded from the analysis because it lies above $T_c^{\text{Ising}}\approx2.269$ and therefore has no ordered phase, and (as expected) we see paramagnetism here at all $p$.

Error bars on $(p_c,\nu,\nu z)$ add in quadrature a statistical term (a bootstrap over the in-window pairs, with $500$ refits) and a systematic term from varying the fit window (refitting as the included range of $p$ is changed). The systematic term dominates; the same approach is used for the $T=0$ fits in Fig.~2 of the main text. Where too few points remain to estimate it (the highest-$T$ $\tau$ fits), the error is reported as undefined rather than zero.

\begin{figure}[h!]
    \centering
    \includegraphics[width=0.85\textwidth]{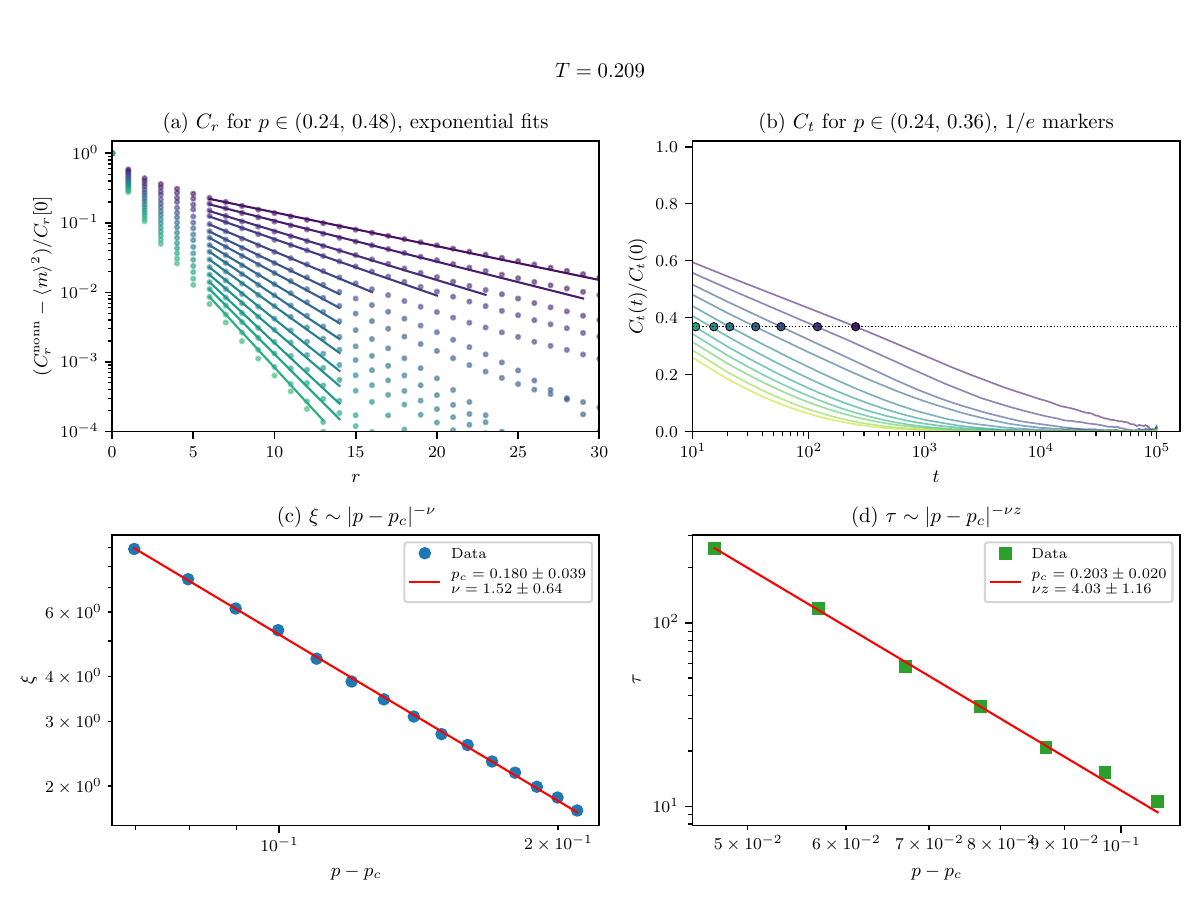}\\
    \caption{\label{fig:sm_fits_1} Per-temperature extraction at $T=0.209$, far below the critical temperature of the clean Ising model $T_c^{\text{Ising}}$ ($p_\text{dem}=0.240$). In each block: (a)~normalised connected spatial correlator $C_r(r)$ with exponential fits; (b)~autocorrelation $C_t(t)$ with $1/e$ markers defining $\tau$; (c)~divergence of $\xi$ with fitted power law [Eq.~\eqref{eq:sm_powerlaws}]; (d)~divergence of $\tau$. The extracted $p_c$, $\nu$ and $\nu z$ are quoted in the legends.}
\end{figure}

\begin{figure}[p]
    \centering
    \includegraphics[width=0.76\textwidth]{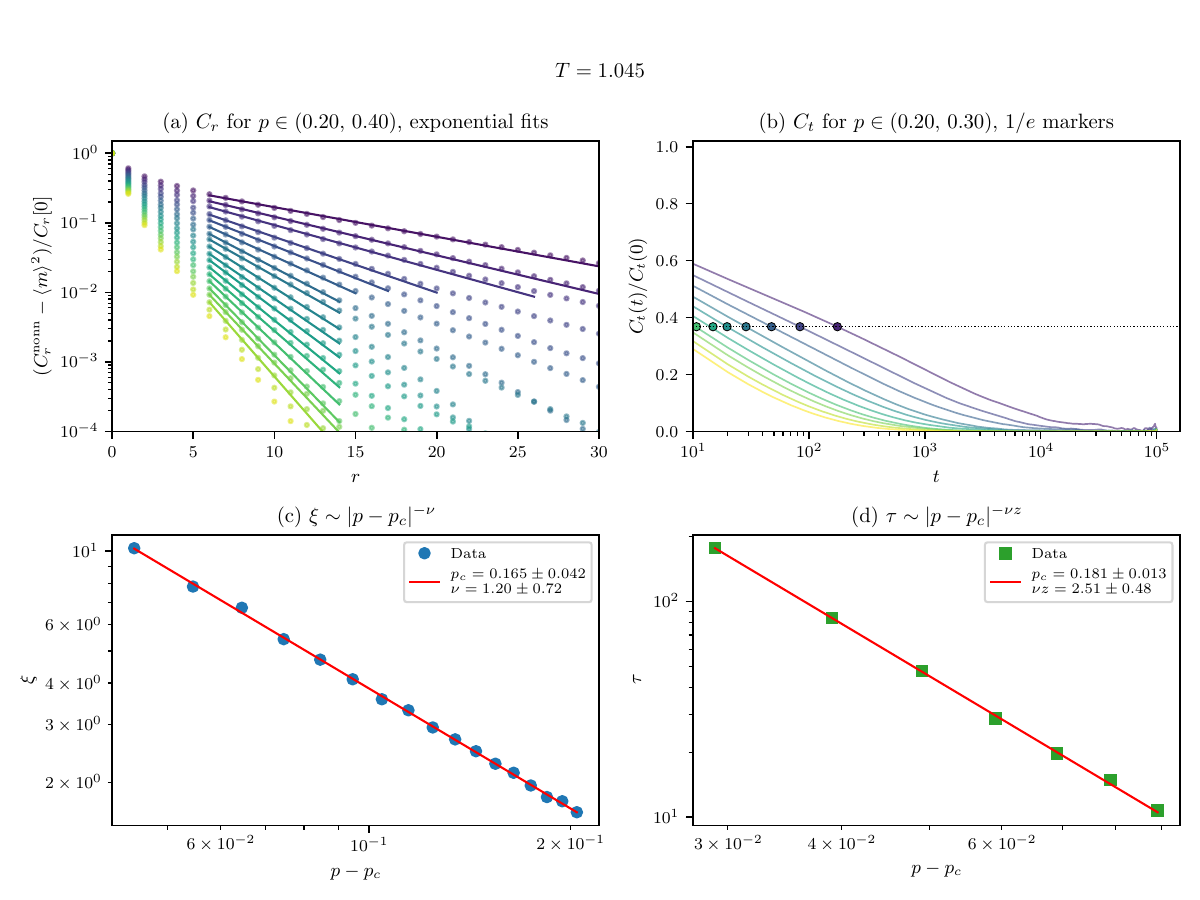}
    \caption{\label{fig:sm_fits_2} Per-temperature extraction at $T=1.045$, i.e., an intermediate temperature between $0$ and $T_c^\text{Ising}$ ($p_\text{dem}=0.200$). Same panels as in Fig.~\ref{fig:sm_fits_1}.}
\end{figure}

\begin{figure}[p]
    \centering
    \includegraphics[width=0.76\textwidth]{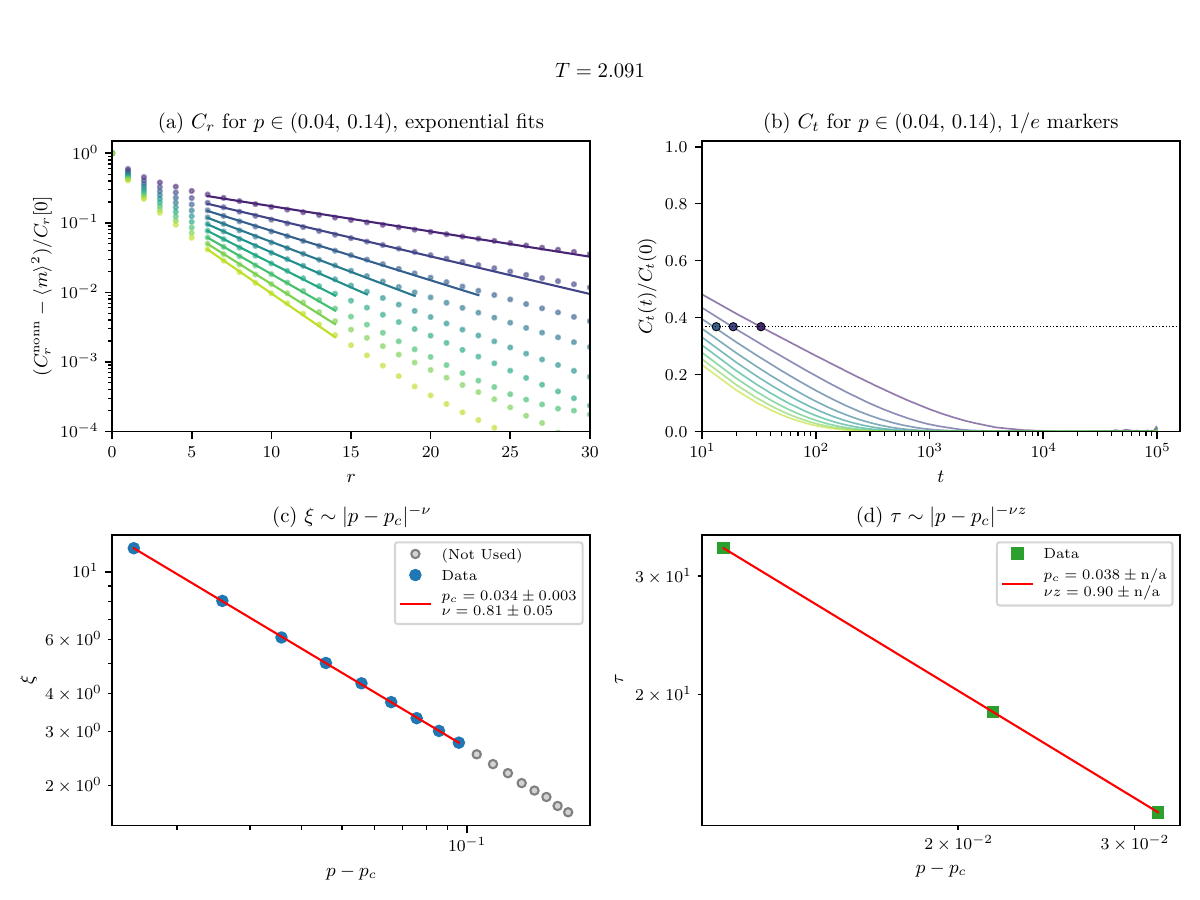}
    \caption{\label{fig:sm_fits_3} Per-temperature extraction at $T=2.091$, just below $T_c^{\text{Ising}}$ ($p_\text{dem}=0.040$). Panels as in Fig.~\ref{fig:sm_fits_1}. Grey points are disordered runs outside the fit window; ordered runs ($p<p_{\mathrm{dem}}$) are excluded from all panels. Near the clean Ising transition the fit window narrows and few points survive (especially for $\tau$), so the exponents extracted here are under-resolved (see text).}
\end{figure}

Figures~\ref{fig:sm_fits_1}, \ref{fig:sm_fits_2}, and \ref{fig:sm_fits_3} show the full extraction at three representative temperatures: $T\approx0.21$ (for the clean Ising model, this is deep inside the ferromagnetic phase), $T\approx1.04$ (intermediate), and $T\approx2.09$ (just below $T_c^{\text{Ising}}$). Each four-panel block contains (a) the normalised connected spatial correlator $C_r(r)$ for the disordered $p$ values (colours), with the exponential fits of Eq.~\eqref{eq:sm_xi_fit} overplotted; (b) the autocorrelation $C_t(t)$ with markers at the $1/e$ crossing defining $\tau$; (c) the divergence of $\xi$ with the fitted power law; and (d) the same for $\tau$. Grey points in Fig.~\ref{fig:sm_fits_3}(c) are disordered runs outside the fit window; ordered runs ($p<p_{\mathrm{dem}}$) are excluded from all panels.

\clearpage

\subsection{Phase diagram and critical exponents}

\begin{figure}[h]
    \centering
    \includegraphics[width=0.46\textwidth]{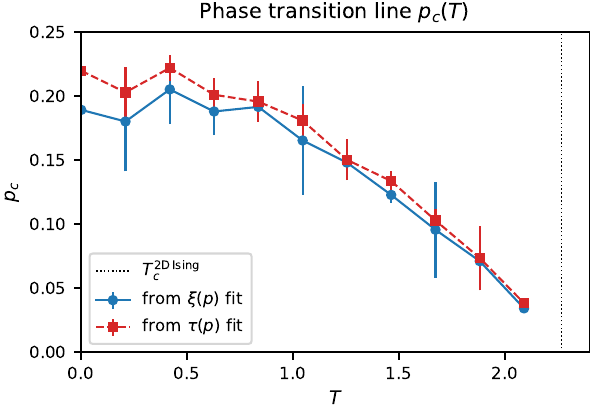}\\[0.6em]
    \includegraphics[width=0.46\textwidth]{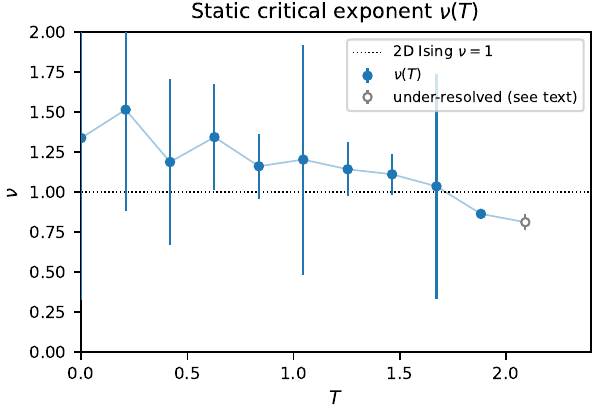}\hfill
    \includegraphics[width=0.46\textwidth]{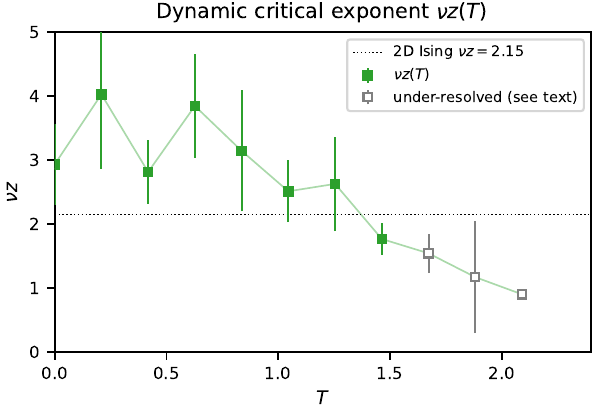}
    \caption{\label{fig:sm_summary} Top: transition line $p_c(T)$, extracted independently from the divergences of $\xi(p)$ (blue) and $\tau(p)$ (red); the dotted vertical line marks $T_c^{\text{Ising}}\approx2.269$. Bottom left: static critical exponent $\nu(T)$ from the divergence of $\xi$; the dotted line is the $2d$ Ising value $\nu=1$. Bottom right: dynamic combination $\nu z(T)$ from the divergence of $\tau$. Error bars are bootstrap $1\sigma$ estimates.}
\end{figure}

Repeating the extraction across all $T$ slices yields the transition line $p_c(T)$ and the estimated exponents $\nu(T)$, $\nu z(T)$ shown in Fig.~\ref{fig:sm_summary}. The two $p_c$ estimators (from $\xi$ and from $\tau$) agree within the quoted errors across the explored range, giving an internal cross-check on the location of the transition. The transition line interpolates between the zero-temperature point $p_c(0)\approx0.20$ and $T_c^{\text{Ising}}\approx2.269$ at $p=0$.

The exponents $\nu(T)$ and $\nu z(T)$ vary, with both seemingly larger at low $T$. At high temperatures, the accessible fits have $p-p_c \sim p_c$ rather than the asymptotic $p-p_c \ll p_c$, and so we cannot expect the extracted exponents to be quantitatively accurate (a more appropriate approach for extracting the high-temperature exponents would be to scan along $T$ at fixed $p$).

Note that the zero-temperature fits as shown in Fig.~1(a) of the main text are not the same as those extracted from the data in Fig.~2; the zero-temperature data in Figs.~1(c), 2, and 3 come from separate, larger simulations with more disorder samples and longer run times, to resolve the critical behaviour and the long-time tails at $T=0$.

\clearpage
\subsection{Finite-time scaling collapse}
\begin{figure}[h]
    \centering
    \includegraphics[width=0.95\textwidth]{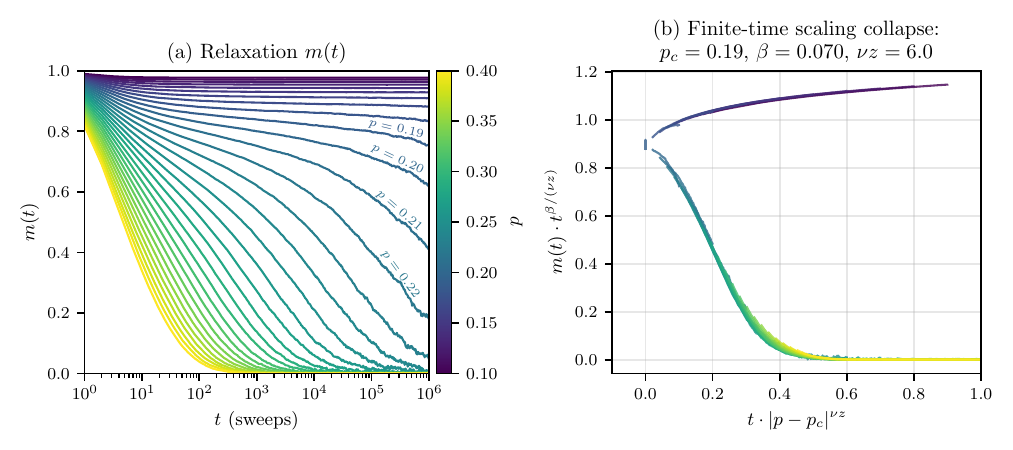}
    \caption{\label{fig:scaling_collapse} (a) Magnetisation $m(t)$ vs time $t$ from an ordered ($m_0=1$) initial condition, for 31 values of $p \in [0.1, 0.4]$ at $T=0$. Each $p$ line is an average over $64$ realisations of disorder in a $512^2$ lattice. (b) Finite-time scaling collapse of the same data at non-early times $t\in[10^2, 10^6]$, using the functional forms of Eq.~\eqref{eq:sm_fts} and parameters of Eq.~\eqref{eq:sm_fts_values}. This collapse was obtained through visual fitting; an estimate of the error in the fitted exponents is not available. }
\end{figure}

The continuous nature of the zero-temperature transition can be confirmed through a finite-time scaling collapse of $m(t)$, relaxing from an ordered initial condition ($m_0=1$). Near a continuous transition, a dynamic scaling ansatz (valid for RBIM at moderate and late times) for relaxation from an ordered initial condition  reads~\cite{Zheng_1998,Luo_Schulke_Zheng_2001}
\begin{equation}\label{eq:sm_fts}
    m(t,p) = t^{-\beta/\nu z}\,\mathcal{G}\!\left[(p - p_c)\,t^{1/\nu z}\right].
\end{equation}
At criticality ($p=p_c$), one expects the magnetisation to decay as a pure power law $m\sim t^{-\beta/\nu z}$; for $p<p_c$ the curves saturate to a finite value in the ordered phase, and for $p>p_c$ decay faster than power law in the disordered phase. Plotting $t^{\beta/\nu z}\,m(t)$ against $\abs{p-p_c}\,t^{1/\nu z}$ collapses the relaxation curves at different $p$ onto a single master curve $\mathcal{G}$, fixing $p_c$, $\nu z$, and $\beta$. We are not aware of any theoretical work discussing the validity of such finite-time scaling forms in nonreciprocal systems, but we obtain a good collapse [Fig.~\ref{fig:scaling_collapse}(b)] with
\begin{equation}\label{eq:sm_fts_values}
    p_c = 0.19, \qquad \nu z = 6.0, \qquad \beta = 0.070.
\end{equation}
The critical $p_c$ agrees with the independent estimates from the divergences of $\xi$ and $\tau$ ($p_c\simeq0.20$, Fig.~2 of the main text). The product $\nu z = 6.0$ is somewhat larger than the value $\nu z\approx4.1$ obtained from the divergence of the bulk relaxation time; we attribute this to finite-size and finite-time corrections to scaling, which are known to be severe for dynamic exponents in disordered systems (see, e.g., Ref.~\cite{Luo_Schulke_Zheng_2001}), rather than to any inconsistency in $p_c$. Further work is required to obtain quantitatively precise scaling exponents.

The order-parameter exponent $\beta=0.070$ is small, well below the clean $2d$ Ising value $\beta=1/8$. However, this effect is common in low-temperature phase transitions of disordered systems; in RBIM, for instance, $\beta=0.095(5)$~\cite{ParisenToldin_Pelissetto_Vicari_2009} (and in RFIM $\beta\approx0.017$~\cite{Middleton_Fisher_2002}). Again, a quantitatively accurate estimate of $\beta$ would require more careful numerical investigation and finite-size scaling analysis.

\clearpage
\section{Nonequilibrium rare-region theory of autocorrelations}
\label{sm:rare}

In the disordered phase ($p>p_c$, including $T=0$) the bulk relaxes on a finite timescale $\tau$, but the \emph{tails} of $C_t(t)$ decay slower than a regular exponential. We attribute this to rare regions: patches in which the local NR bond density is anomalously low, which order locally and reverse only on a timescale $\gg\tau$. This is the $T=0$, athermal analogue of the rare-region mechanism in equilibrium disordered magnets~\cite{Bray1988, Vojta_2006}; the key difference is that here the rare-region reversals are driven by NR defects (as well as thermal noise), so they persist down to $T=0$ where an equilibrium system would freeze.

Consider a $2d$ patch of linear size $\ell$. Since each bond is independently NR with probability $p$, the probability of an anomalously ``clean'' patch (local density $p'<p_c(T)$, low enough to order) scales with its \emph{area},
\begin{equation}\label{eq:sm_Pl}
    P(\ell) \sim \exp[-c(p)\,\ell^{2}],
\end{equation}
where $c(p)>0$ is the cost per unit area. Since magnetisation of a rare region requires the local $p'< p_c(T)$ (ignoring finite-size corrections to $p_c$), this becomes more likely as $p \to p_c^+$ (as a smaller local fluctuation in $p$ is required), and so $c(p\to p_c^+)\to0$~\cite{Vojta_2006}. We take the mean reversal time of a region to grow as a power of its size,
\begin{equation}\label{eq:sm_taul}
    \tau_\ell \sim \ell^{\,\zeta}, \qquad \zeta>0.
\end{equation}
This is an ansatz (we are not aware of out-of-equilibrium predictions for $\tau_\ell$ in this context), but reproduces the observed tails.

A patch surviving to time $t$ contributes coherently to $C_t(t)$. Weighting each size by its density and survival probability $e^{-t/\tau_\ell}$ yields
\begin{equation}\label{eq:sm_Ct_integral}
    C_t(t) \sim \int_0^\infty d\ell\; e^{-c(p)\,\ell^{2} - t\,\ell^{-\zeta}}.
\end{equation}
Large patches are rarer but longer-lived than smaller patches; these two effects compete, and for large $t$ the integral is dominated by a saddle at $\ell^*(t)$. Extremising the exponent $\Phi(\ell)=c(p)\,\ell^2 + t\,\ell^{-\zeta}$ gives $\ell^*(t)\sim \{\zeta t/[2\,c(p)]\}^{1/(\zeta+2)}\sim t^{1/(\zeta+2)}$, at which both terms scale as $t^{2/(\zeta+2)}$, so
\begin{equation}\label{eq:sm_stretched}
    C_t(t) \sim \exp[-(t/\tau^{\mathrm{(str)}})^{\alpha}],
    \qquad \alpha = \frac{2}{\zeta+2}<1,
\end{equation}
with $\tau^{\mathrm{(str)}} \sim c(p)^{-\zeta/2}$. Two predictions follow, which are consistent with the numerical observations in Fig.~3(b2) of the main text: $\tau^{\mathrm{(str)}}$ diverges as $p\to p_c^+$ [since $c(p)\to0$], and $\alpha$ is fixed by $\zeta$, approaching a constant at criticality.

\paragraph{Crossover at large $p$.} The saddle point assumes $\ell^*(t)$ is large compared to the lattice spacing. As $p$ grows away from $p_c$, $c(p)$ increases and $\ell^*$ shrinks to $\mathcal{O}(1)$: the continuum forms \eqref{eq:sm_Pl}--\eqref{eq:sm_taul} break down, and so well-defined rare regions disappear, reverting the autocorrelation tail to a simple exponential ($\alpha\to1$). This matches the increase of fitted $\alpha$ towards unity with $p$ in Fig.~3(b2).

\paragraph{Spatial correlations are unaffected.} Rare regions dominate the tail of $C_t(t)$ but leave $C_r(r)\sim e^{-r/\xi}$ unaffected [Fig.~1(c1)]. The reason is how each correlator samples them: $C_r(r)$ counts equal-time pairs sharing a patch, which requires $\ell\gtrsim r$, with weight
\begin{equation}\label{eq:sm_rarefraction}
    w_{\mathrm{rare}}(r) \sim \int_r^\infty d\ell\;\ell\,e^{-c(p)\ell^2} \sim e^{-c(p)\,r^2}.
\end{equation}
This is \emph{Gaussian} in $r$, faster than the bulk $e^{-r/\xi}$, so the rare-region piece is a subdominant additive correction and the measured $\xi$ is the bulk length.

\end{document}